\documentclass[ jmp, amsmath,amssymb,reprint, onecolumn]{revtex4-1}
\usepackage{graphicx}
\usepackage{dcolumn}
\usepackage{bm}
\usepackage{color}

\usepackage[utf8]{inputenc}
\usepackage[T1]{fontenc}
\usepackage{mathptmx}
\usepackage{etoolbox}
\usepackage{subfigure}
\def\I{{\mathbb I}}

\def\H{{\mathcal H}}

\def\N{{\mathbb N}}
\def\R{{\mathbb R}}

\def\vm{{\mathbf m}}
\def\vn{{\mathbf n}}
\def\va{{\mathbf a}}
\def\vb{{\mathbf b}}
\def\vx{{\mathbf x}}

\newcommand{\ds}{\displaystyle}

\makeatletter
\def\@email#1#2{
 \endgroup
 \patchcmd{\titleblock@produce}
  {\frontmatter@RRAPformat}
  {\frontmatter@RRAPformat{\produce@RRAP{*#1\href{mailto:#2}{#2}}}\frontmatter@RRAPformat}
  {}{}
}
\makeatother
\begin{document}

\preprint{AIP/123-QED}

\title[Sample title]{Gelfand Triplets, Continuous and Discrete Bases and Legendre Polynomials}

\author{E. Celeghini$^1$} 
\author{M. Gadella$^2$}
\author{M. A. del Olmo$^2$}

\affiliation{ $^1$ Dipartimento di Fisica, Universit\`a di Firenze and INFN-Sezione di Firenze, 
  {\bf I}50019 Sesto Fiorentino, Firenze, Italy\\
  $^2$ Departamento de F\'{\i}sica Te\'orica, At\'omica y Optica  and IMUVA, Universidad de Va\-lladolid, 47011 Valladolid, Spain}

\email{enrico.celeghini@gmail.com, manuel.gadella26@gmail.com, marianoantonio.olmo@uva.es}

\date{\today}

\begin{abstract}
We consider  a  basis of square integrable functions on a rectangle, contained in $\R^2$, constructed with Legendre polynomials, suitable,  for instance, for  the analogical description of images on the plane or in other fields of application of the Legendre polynomials in higher dimensions. 
  After extending the Legendre polynomials to any arbitrary interval of the form $[a,b]$, from its original form on $[-1,1]$, we generalize the basis of
  Legendre polynomials to two dimensions. This is the first step to generalize the basis to $n$-dimensions.
  We present some mathematical constructions such as  Gelfand triples appropriate on this context. ``Smoothness'' of functions on space of test functions and some other properties are revisited, as well as te continuity of generators of $su(1,1)$ on this context.

\end{abstract}

\maketitle

\section{Introduction and Motivation}\label{intro}

Orthogonal polynomials and their generalizations are being widely used in many areas of the broad field of signal processing as we may see, for instance, in
Ref.~\cite{QZWZC23,JRM12,see2007,zhoushutoumoulin2005,huntmukundam2004,paramersranong2003,mukundamong2001,teague1980, hu1962}. In particular \cite{QZWZC23} contains an extensive list of additional references. 
Since the seminal
paper by Teague in 1980 \cite{teague1980}, Legendre polynomials via Legendre moments have also been largely used \cite{mukundanramakrishnan1995,yapparamesran2005,chongraveendranmukundan2004,xiaowangli2014,YangMaMiaoLiuWangMeng2019,hosnydarwishaboelenen2020}. 
  
As we know, we require a normalization in order to use a Hilbert space basis with Legendre polynomials  \cite{szego2003}. This normalization is given by a multiplicative factor
on the Legendre polynomials $\{L_n(x)\}_{n\in \N}\,$ given by   $\sqrt{n+1/2}$  \cite{olmo2013}. By historical
reasons related to trigonometry, Legendre polynomials are usually defined in the interval\, $[-1,+1]$. In fact, the original problem discussed by Legendre, which gave rise to these kind of polynomials showed spherical symmetry   \cite{legendre1785}. 

The scientific discussion contained in the present article has a motivation with roots on the signal theory and the theory of images \cite{celeghini2017}. As the ultimate purpose would be the use of
the above mentioned basis in the description of a set of data of a digital sample, it would be necessary to define these functions on a general compact interval such as $[a,b]$ (with $a<b$),
supporting the original data of the problem to be discussed. 

The natural definition of Legendre polynomials on intervals of the form $[a,b]\subset \R$ are valid for data analysis in one dimension (1D). However, for 2D or 3D, we need to extend
the notion of Legendre polynomials to these contexts. In these cases, as well as  in the analysis of data from images, we need bases of functions supported on a 2D or 3D real subset, which are constructed    using Legendre
polynomials and are addressed to the study of black and white images. In the case of color images one needs to work in a higher dimensions,   as discussed in the above mentioned references. This motivates the use of mathematical tools such as bases of functions in 2D built with  Legendre polynomials, which are the
normalized versions of $\{L_m(x) \otimes L_n(y)\}_{m,n\in \N}$, with $x\in [a_x,b_x]\subset \R$ and $y\in [a_y,b_y]\subset \R$.

This is just about the physical motivation of the mathematical discussion that we carry along the present manuscript. From this point of view, we are taking into account several issues. First of all, we go through a detailed description of the 1D Legendre polynomials on a real interval
of the form $[a,b]$.  Particular interest are those for which $a=0$, since the construction is related to the standard discrete numbering in computer science where the numbering of
pixels starts from pixel\;${1}$ and finishes with pixel  ${n}$. Nevertheless, the use of intervals of the form $[a,b]$ have their applications to discrete image analysis,
when one just considers a part of pixels in one given image.

In a second step, we intend to extend the analysis in 1D to 2D rectangular images on sets of the form $[a_1,b_1] \otimes [a_2,b_2]\subset \R^2$, where the involved parameters are, in principle, arbitrary albeit
finite. Physically, this is related to the description of a rectangular black and white image, where the saturation of the black at each point $(x,y)$ is given by
\begin{equation}\label{1}
f(x,y)\,  =\, \sum_{n_1,n_2} \; d_{n_1,n_2}\;\; W_{n_1}(a_x,b_x,x) \otimes W_{n_2}(a_y,b_y,y)\,.
\end{equation}
Here,  $\{W_n(a,b,x)\}$ are the Legendre polynomials considered as functions supported on a compact interval of the form $[a,b]$.
Consequently, the tensor products $W_{n_1}(a_x,b_x,x) \otimes W_{n_2}(a_y,b_y,y)$ give the extension of these functions to the 2D rectangle $[a_x,b_x] \otimes [a_y,b_y]$. 
Some of these ideas has
been proposed in \cite{olmo2019,CGO22}.

The central structure accommodating the functions for rectangular black and white images is the Hilbert space $L^2([a_x,b_x] \otimes [a_y,b_y])$. Nevertheless, equipations of Hilbert spaces such as Gelfand triplets \cite{GEL}  are
often important so as to provide subspaces in which observables are continuous operators. Gelfand triplets or rigged Hilbert spaces (RHS) are a sequence of three infinite dimensional vector spaces,
\begin{equation}\label{2}
\Phi \subset \mathcal H \subset \Phi^\times\,,
\end{equation}
where $\mathcal H$ is a separable infinite dimensional complex Hilbert space, $\Phi$ is a dense subspace endowed with a locally convex topology finer than the Hilbert space topology.
Finally, $\Phi^\times$ is the dual space of $\Phi$, usually endowed with a topology compatible with the dual pair $\{\Phi,\Phi^\times\}$, normally the strong or weak topology. Gelfand
triplets has many applications either in mathematics or theoretical physics. Due to the wide range of these applications, we only mention here the mathematical formalization of the
Dirac formalism of Quantum Mechanics \cite{ROB,ANT,MEL,BOH,GG}, the signal processing \cite{olmo2016} and others \cite{TTT}.

This paper is organized as follows: In Section ~\ref{realline}, we revise some features on orthogonal polynomials and their use as basis of square integrable functions on the real line or real
intervals \cite{olmo2016}. In Section~\ref{legendre}, we derive the definition of Legendre polynomials on an arbitrary compact interval $[a,b], a<b$, starting from the usual definition on $[-1,1]$.
These are the so called generalized Legendre polynomials,  $W_n(a,b,x)\equiv W^{[a,b]}_n(x)$. We show that they have similar properties than the standard Legendre polynomials $L_n(x)$.
In Section~\ref{su11}  we show that the generalized Legendre polynomials $W^{[a,b]}_n(x)$, in complete analogy with the usual Legendre polynomials $L_n(x)$, for any pair of real numbers
$a,b$  $(a<b)$ are a unitary irreducible representation (UIR) of the group $SU(1,1)$ with Casimir ${\cal C}=-1/4$. A generalization of Legendre polynomials on n-dimensional (nD)
rectangles appears in Section~\ref{nlegendre}. In Section VI, we construct Gelfand triplets spanned by the generalized Legendre polynomials, analyze the derivability of the functions on the test space and define a sort of continuous basis for functions in the test space. In Section VII, we prove the continuity of the generators of $su(1,1)$, and therefore of all enveloping algebra, on the test space.
The paper closes with few concluding remarks.

\section{Bases on the Real Line and Orthogonal Polynomials}\label{polynomials}\label{realline}

To begin with, let us consider orthogonal polynomials defined on real intervals of the form $[a,b]$, either finite or infinite, i.e., $-\infty \le a < b \le +\infty$ \cite{cambanis}.
One standard procedure described in Quantum Mechanics textbooks comes after the association of, in general almost all $x\in [a,b]$ with respect to some measure, a generalized vector,
$|x \rangle$, outside the Hilbert space, although in the dual $\Phi^\times$ of a Gelfand triplet (RHS) of the form given in \eqref{2}. Constructions of this kind come after the
celebrated Gelfand-Maurin theorem \cite{GEL,ROB,ANT,MEL,GG1,GG2}, which we do not discuss it in here. These generalized vectors form a generalized basis, which plays an interesting
role on the discussion in the sequel.

In most of examples studied so far \cite{olmo2016,olmo2019,CGO22,CGO20}, we have the similar structure: an infinite dimensional separable Hilbert space $\H_{E_{a,b}}$  equipped as in \eqref{2}, where
$E_{a,b}:= [a,b] \subset \R$. To all (or almost all with respect to some measure) $x\in [a,b]$, it corresponds a generalized vector on the dual,
$|x\rangle$ (related to $\langle x| \in \Phi^\times$) see \eqref{2}, fulfilling some formal properties of orthogonality and completeness, which are usually presented as, respectively,
\begin{equation}\label{3}
\langle x | y \rangle = \delta(x-y)\,,  \qquad  \int_a^b |x\rangle dx \langle x| = \I\,, \qquad 
\quad\; \forall \;x,y\; \in \,E_{a,b}\,.
\end{equation}

The abstract Hilbert space admits a representation ${L}^2(E_{a,b})$, where an orthonormal basis (complete orthonormal set) is a sequence of functions. In our examples, we
have chosen bases of special functions $\{F_n(x)\}_{n\in \N}$ for which the orthonormality and completeness relations read, respectively, as

\begin{equation}\label{4}
\int_{a}^{b} \, F_n(x)\; F_m(x)\; dx\;=\,\delta_{n,m} \,,\qquad
\sum_{n=0}^\infty \;F_n(x)\; F_n(y) \;=\; \delta(x-y)\,,\quad\; \forall\; x, y \in E_{a,b} \,.
\end{equation}

In each of our examples, we have constructed a Gelfand triple of the form
\begin{equation}\label{5}
\Phi_{E_{a,b}} \subset L^2(E_{a,b}) \subset \Phi^\times_{E_{a,b}}\,,
\end{equation}
such that $F_n(x) \in \Phi_{E_{a,b}}$. We have shown that there exists a unitary operator, $U:\H_{E_{a,b}} \longmapsto {L}^2(E_{a,b})$, not necessarily unique, such
that if $\Phi=U\Phi_{E_{a,b}}$, then
\begin{equation}\label{6}
\Phi \subset \H_{E_{a,b}} \subset \Phi^\times\,,
\end{equation}
is an abstract Gelfand triple with the property that $\{|x\rangle\}_{x \in E_{a,b}} \subset \Phi^\times_{E_{a,b}}$. 
Hereafter, we construct an orthonormal basis, $\{|n\rangle\}_{n\in\N}$ on the abstract Hilbert space $\H_{E_{a,b}}$ 
 such that  $|n\rangle := UF_n$. It is shown that it satisfies the formal expression
 \begin{equation}\label{7}
 |n\rangle\, := \int_{a}^{b} \; |x\rangle \; F_n(x) \; dx\,.
\end{equation}
Orthonormality and completeness relations for $\{|n\rangle\}_{n\in\N}$ are given by the following respective relations:
\begin{equation}\label{8}
 \langle n|m \rangle \;=\; \delta_{n,m}\,,\qquad  \sum_{n=0}^\infty \; |n\rangle \langle n|\;=\;\I \,,  \quad \forall\; n,m \in \N\,,
\end{equation}
where $\mathbb I$ is the identity on $\H_{E_{a,b}}$.

In terms of the generalized vectors $\{|x\rangle\}_{x \in E_{a,b}}$ and the orthonormal basis $\{|n\rangle\}_{n\in\N}$, we may write the functions $\{F_n(x)\}_{n\in \N}$, assumed to be real, as
\begin{equation}\label{9}
F_n(x)\;:= \;\langle x|n \rangle \;=\; \langle n|x \rangle\,.
\end{equation}

An arbitrary vector $|f\rangle \in \Phi_{E_{a,b}}$ can also be expressed in terms of both, the orthonormal basis $\{|n\rangle\}_{n\in\N}$ as well as in the set of generalized
vectors $\{|x\rangle\}_{x \in E_{a,b}}$ in the following way:
\begin{equation}\label{10}
|f\rangle \;=\; \int_{a}^{b}\; |x\rangle \,dx\, \langle x|f\rangle \;\;=\;
\sum_{n=0}^{\infty} \; |n\rangle \; \langle n|f\rangle\,.
\end{equation}
From this point of view, we may assert that the set $\{|x\rangle\}_{x \in [a,b]}$ form a {\it generalized continuous basis} for all vectors $|f\rangle \in \Phi_{E_{a,b}}$. Also,
any function $f(x) \in \Phi$, with $f(x) =U|f\rangle$ can be written as follows:
\begin{equation}\label{11}
f(x)= \;\langle x|f \rangle \, =\,
\sum_{n=0}^{\infty} f_n \;F_n(x) \,,\qquad f_n\,=\, \langle n|f\rangle \,=\,
\int_{a}^{b}\, f(x)\;F_n(x)\;dx\,. 
\end{equation}
Obviously, the $\{f_n\}_{n\in\N}$ is the set of coefficients of the span of any $f(x) \in \Phi$ in terms of the given orthonormal basis. This means that
\begin{equation}\label{12}
\langle g|f \rangle \;=\, \sum_{n=0}^{\infty}\; g_n^* \cdot f_n \;=\; \int_{a}^{b} g^*(x) \cdot f(x) dx\,,
\qquad \sum_{n=0}^{\infty} |f_n|^2 \;=\, \int_{a}^{b} |f(x)|^2 dx\,.
\end{equation}

As we have remarked earlier, we use normalized orthogonal polynomials as   orthonormal bases on $L^2(E_{a,b})$. In this sense, we have shown that Gelfand triplets or RHS are the
suitable setting that give meaning to formulas including special functions, generalized vectors, generalized eigenvalues of operators and discrete orthonormal and continuous
bases. Moreover, it allows for representations of Lie algebras of symmetries of physical systems, as {\it continuous operators} on a locally convex space contained on their
domain. Needless to say that the type of basis of special functions depends, among other considerations, of the kind of interval $[a,b]$ used. Here, there are some
examples \cite{szego2003,OLBC2010,olmo2013}:

2.1) Assume that $a=-\infty$, $b=+\infty$, i.e., $-\infty < x < \infty$. The Hilbert space ${L}^2(E_{a,b})$ is $L^2(\R)$. An orthonormal basis on $L^2(\R)$ is given by the
{\it normalized Hermite functions}, $\{K_n(x)\}_{n\in \N}$, defined as
\begin{equation}\label{13}
 K_n(x) :=\;\; \frac{e^{-x^2/2}}{\sqrt{2^n \, n!\sqrt{\pi}}}\;\; H_n(x) \,,\qquad n \in \N\,, \qquad n=0,1,2,\dots\,,
\end{equation}
where $\{H_n(x)\}_{n\in \N}$ are the Hermite polynomials. For an unbounded interval, our special functions are products of some polynomials times a regularization
function, in our case a gaussian, times a normalization constant. 
A typical equipation of $L^2(\R)$ is given by $\mathcal S \subset L^2(\R) \subset \mathcal S^\times$, where $\mathcal S$ is the Schwartz space  \cite{simon1980}, which
contains all Hermite functions, and $\mathcal S^\times$ is the space of tempered distributions.

2.2) Now, take, $a=0$ and $b=+\infty$, so that ${L}^2(E_{a,b}) \equiv L^2(\R^+)$.  An orthonormal basis is given by the {\it Laguerre functions}, which are products of
Laguerre polynomials times a negative exponential:
\begin{equation}\label{14}
M_n(x):= e^{-x/2}\; {L}_n(x), \qquad n \in \N \,,
  \; x \in [0,+\infty)\equiv \R^+\,.
\end{equation}

2.3) If $a=-1$ and $b=1$, we have the Hilbert space $L^2[-1,1]$, where the {\it normalized Legendre polynomials} form an orthonormal basis.

\section{Legendre polynomials and their generalization}\label{legendre}

The Legendre polynomials,

\begin{equation}\label{15}
\ P_n(x) := \frac{1}{2^n\,n!}\, \frac{d^n}{dx^n} \,(x^2-1)^n\,,
\end{equation}
are defined on the interval $[-1,1]$ and verify the following orthonormality and completeness relations:
\begin{equation}\label{16}
\int_{-1}^{1} \,  P_n(x)\,(n+1/2)\,  P_m(x)\; dx\;=\,\delta_{n,m} \,,\qquad
\sum_{n=0}^\infty \; P_n(x)\,(n+1/2)\,  P_n(y) \;=\; \delta(x-y)\,.
\end{equation}
This means that the normalized Legendre polynomials (NLP) are 
\begin{equation}\label{17}
K_n(x):= \sqrt{n+1/2}\,\ P_n(x)\,.
\end{equation}
 They
 are an orthonormal basis on $L^2[-1,1]$, the space of measurable square integrable complex functions on $[-1,1]$. From this, we may construct an orthonormal basis  for $L^2[a,b]$  in terms of the functions  $W_n(a,b,x)$, with $a$ and $b$ finite, as follows:
\begin{equation}\label{18} 
W_n(a,b,x) := \sqrt{\frac{2}{b-a}} \;\;\; K_n \left(\frac{2x}{b-a}-\frac{b+a}{b-a}\right)\,,
\qquad\; n \in \N\,, \;\; a \le x \le b\, .
\end{equation}
When $b-a=1-(-1)$, i.e., the interval $[a,b]$ has the same width than $[-1,1]$, the argument of $P_n(x)$ reduces to $x-(b+a)/2$. Functions $W_n(a,b,x)$ are the {\it generalized Legendre polynomials} (GLP) on the interval $[a,b]$. 

\begin{figure}[h]
\centering
 \subfigure [$K_n(x)$ \hskip6.95cm (b)   $ W_n(3,7,x)$]{\includegraphics[width=0.45\textwidth]{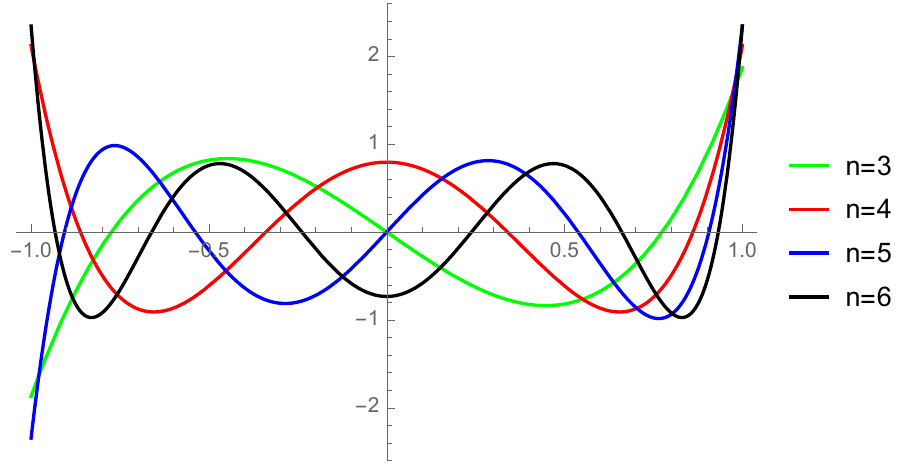}\qquad  
\includegraphics[width=0.45\textwidth]{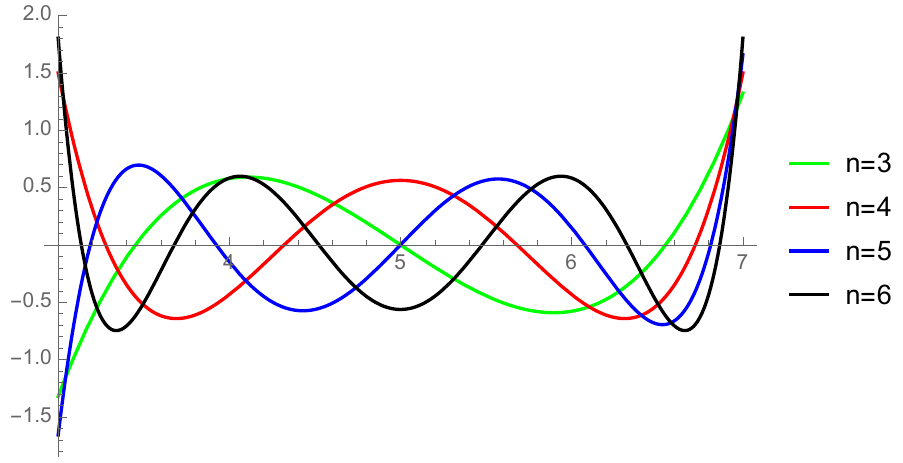}}
\caption{\footnotesize Graphics of  (a) the normalized Legendre polynomials  $K_n(x)$  \eqref{18} in their usual interval $[-1,1]$ and (b) the generalized  Legendre polynomials  $W_n(a,b,x)$ \eqref{19} defined in the interval $[3,7]$ for different values of $n$.}
\label{figure_1}
\end{figure}

In Fig.~\ref{figure_1}.a  the NLP $K_n(x)$ for different values of the label $n$  are represented and in Fig.~\ref{figure_1}.b appear the GLP $W_n(a,b,x)$ defined in the interval $[3,7]$ also for some values of $n$.

From the orthonormality and completeness relations satisfied by the orthonormal Legendre polynomials, we readily obtain similar relations for the new GLP, such that
 \begin{equation}\label{19}
 \begin{array}{lll}
\ds  \int_{a}^{b}  W_n(a,b,x)\, W_m(a,b,x)\, dx  &=&\ds \delta_{n,m}\,,\\[0.4cm]
\ds  \sum_{n=0}^\infty W_n(a,b,x)\, W_n(a,b,y)  &=&\ds \delta(x-y)\,, \quad \forall x,y \in [a,b]\,.
  \end{array}  \end{equation}

Similarly to the recurrence relations which satisfy the Legendre polynomials on $[-1,1]$, $P_n(x)$, (see eqs. (20) of Ref.~\cite{olmo2013}) we have also have recurrence equations for the GLP,  $W_n$, which are:
\begin{equation}\label{20} 
\begin{array}{lll}
 \ds\frac{2}{b-a}  \sqrt{\frac{n+3/2}{n+1/2}}\, \left[(x-a)(x-b)\, W'_n
+(n+1)\left( x-\frac{b+a}{2}\right)\,W_n\right]&=& (n+1)\, W_{n+1}\,, 
 \\[0.4cm]
 \ds   \frac{2}{b-a} \sqrt{\frac{n-1/2}{n+1/2}}\,\left[-(x-a)(x-b)\, W'_n\;
+n \left(x\;-\frac{b+a}{2}\right)\, W_n\right] &=&n\, W_{n-1}\,,
 \end{array}
\end{equation}
where the prime in $W'_n$ ($\equiv W'_n(a,b,x)$) represents derivation with respect to $x$. Obviously, these relations become the usual ones when $a=-1$ and $b=1$.


\section{Legendre polynomials and $SU(1,1)$}\label{su11}

Let us define the following differential operators, $J_\pm$ as
\begin{equation}\label{21} 
J_\pm = \frac{2}{b-a}\,\left[\pm\, (X-a)(X-b) D_x +\left(X-\frac{b+a}{2}\right)\,
\left(N+\frac{1}{2}\pm \frac{1}{2}\right)\
\right]\,
\sqrt{\frac{N+\frac{1}{2}\pm 1}{N+1/2}}\,,
\end{equation}
where the position operator $X$ and the derivative operator $D_x$ act on functions of their own domain as
usual, i.e.,
\begin{equation}\label{22} 
(X\,f)(x)=x\, f(x)\,,\qquad     D_x f(x)=\frac{d\,f(x)}{dx}\,,
\end{equation}
and $N$ is the so called {\it number operator}, acting on the GLP  $W_n(a,b,x)$ as
\begin{equation}\label{23}
N \,W_n(a,b,x)= n\, \,W_n(a,b,x)\,,
\end{equation}

The operators $J_\pm$ act on  $W_n(a,b,x)$  as

\begin{equation}\label{24} 
J_+ \,W_n(a,b,x) = (n+1) \, W_{n+1}(a,b,x)\,,  \qquad   J_- \, W_n(a,b,x) = n \, W_{n-1}(a,b,x)\,.
\end{equation}
Equation \eqref{21} leads to the following pair of diagonal operators:  
\begin{equation}\label{25} 
\begin{array}{lll}
J_+\,J_--N^2  &=&\ds-\frac{4}{(b-a)^2}\,(X-a)(X-b) \\[0.4cm]
&&\ds\;\times \left[-(X-a)(X-b)\,D_x^2\,
-2\left(X-\frac{b+a}{2}\right)\,D_x\;+N(N+1)\right]\,,
 \end{array} 
 \end{equation}
and
\begin{equation}\label{26} 
\begin{array}{lll}
J_-\,J_+ -(N+1)^2  &=&\ds-\frac{4}{(b-a)^2}\,(X-a)(X-b)\\[0.4cm]
&&\ds\;\times\left[-(X-a)(X-b)\,D_x^2\,
-2\left(X-\frac{b+a}{2}\right)\,D_x\;+N(N+1)\right].
\end{array}
\end{equation}
They verify the following relations for all $n\in \N$:
\begin{equation}\label{27} 
\left( J_+\,J_--N^2\right)\, W_{n}(a,b,x)=0\,,\qquad   \left( J_-\,J_+ -(N+1)^2\right)\,W_{n}(a,b,x)=0\,.
\end{equation}
Let us define a new operator  $J_3: = N + 1/2$, so that for any $n \in \N$
\begin{equation}\label{28} 
J_3 \, W_{n}(a,b,x) = (n+1/2)\, W_{n}(a,b,x)\,.
\end{equation}
These operators close the following commutation relations on the linear space spanned by the set of GLP 
$\{W_n(a,b,x)\}_{n\in \N}$:
\begin{equation}\label{29} 
[J_3\,,J_\pm]=\pm\,J_\pm\,, 
\qquad\quad [\,J_+\,,J_-\,]\,=\,-2\,J_3\,.
\end{equation}
These are the commutation relations corresponding to the Lie algebra $su(1,1)$. In addition to the commutation relations \eqref{29}, we have the anticommutation
rule $\{J_+,J_-\}= -2J_3^2 +\frac12$. 

The linear space spanned by the set of GLP,  $\{W_n(a,b,x)\}_{n\in \N}$, for every pair of real numbers $a<b$, supports a unitary irreducible representation
of the group $SU(1,1)$ with quadratic Casimir $\mathcal C=-1/4$, so that
\begin{equation}\label{30}
\left({\mathcal C} \,+ \frac14\right)\, W_n(a,b,x)\equiv \left(J_3^2\,-\frac12\,\{J_+,J_-\}+\frac14\right)\, W_n(a,b,x)=0\,,
\end{equation}
which comes after \eqref{25} and \eqref{26}. Equation \eqref{29} is nothing else that the generalization of the usual Legendre equation on $[-1,1]$ given as
\begin{equation}\label{31}
\left( X\,D_x^2+D_x-\frac14\,X+N+\frac12\right)\, P_n(x)=0\,, 
\end{equation}
to the interval $[a,b]$.


\section{Legendre functions on rectangles of arbitrary dimension}\label{nlegendre}

To begin with, let us take a rectangular domain, $\mathcal R^2$ in $\R^2$:
\begin{equation}\label{32}
 \mathcal R^2=\{(x_1,x_2)\in \R^2\, \big|\, x_1\in [a_1,b_1]\subset \R\,, x_2\in [a_2,b_2]\subset \R\}\,,
\end{equation}
where $a_i$ and $b_i$, $i=1,2$ are finite real numbers and $a_i < b_i$. 
Let us use the following {\it vectorial} notation:
\begin{equation}\label{33} 
\mathcal R^2\equiv [\va,\vb]^2:= [a_1,b_1]\times  [a_2,b_2]\,,\qquad
\va=(a_1,a_2)\,, \quad \vb=(b_1,b_2) \in \R^2\,.
\end{equation}

An orthonormal basis for the space $L^2([\va,\vb]^2)$ is given by the tensor products of the GLP
\begin{equation}\label{34}
W_{\mathbf m}({\bf a},{\bf b},{\bf x}) \equiv W_{m_1}(a_1,b_1,x_1) \otimes W_{m_2}(a_2,b_2,x_2) \,,
\end{equation}
 where ${\mathbf m} = (m_1,m_2)\in \N^2$,  ${\mathbf a} $, ${\mathbf b} \in \R^2$ and ${\mathbf x} \in [\va ,\vb]^2$.  Being the functions \eqref{34} an orthonormal basis, they satisfy orthonormality and completeness relations:
\begin{equation}\label{35}
\begin{array}{lll}
\ds\int_{[\va,\vb]^2} d^{\,2}\vx \, W_{\vm'}({\va},{\vb},{\vx})\,
W_{\vm}({\va},{\vb},{\vx}) =\delta_{\vm,\vm'}\equiv  \delta_{m_1,m_1'}\,\delta_{m_2,m_2'}\,,
\\[3ex]
\ds \sum_{\vm\in \N^2}\, W_{\vm}({\va},{\vb},{\vx})\, W_{\vm}({\va},{\vb},{\vx'}) =\delta (\vx,\vx')=\delta(x_1-x_1')\,\delta(x_2-x_2')\,,\quad
\forall \, \vx, \vx' \in[\va,\vb]^2\,.
\end{array}
\end{equation}  

\begin{figure}[h]
\centering
 \subfigure [$W_3(3,7,x_1)\otimes W_3(3,7,x_2)$ \hskip3.85cm (b)   $W_5(3,7,x_1)\otimes W_7(3,7,x_2)$]{\includegraphics[width=0.4\textwidth]{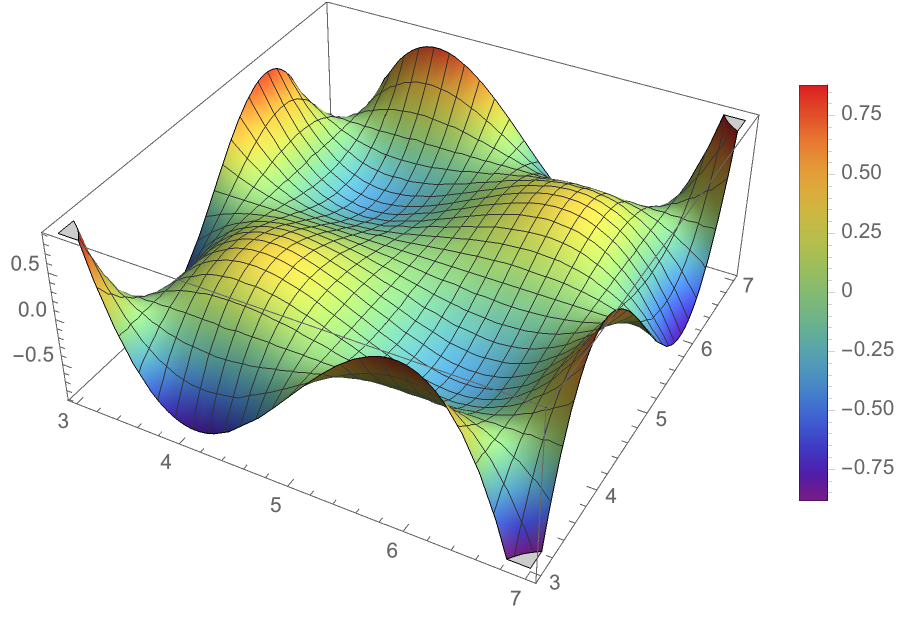}\qquad  
\includegraphics[width=0.4\textwidth]{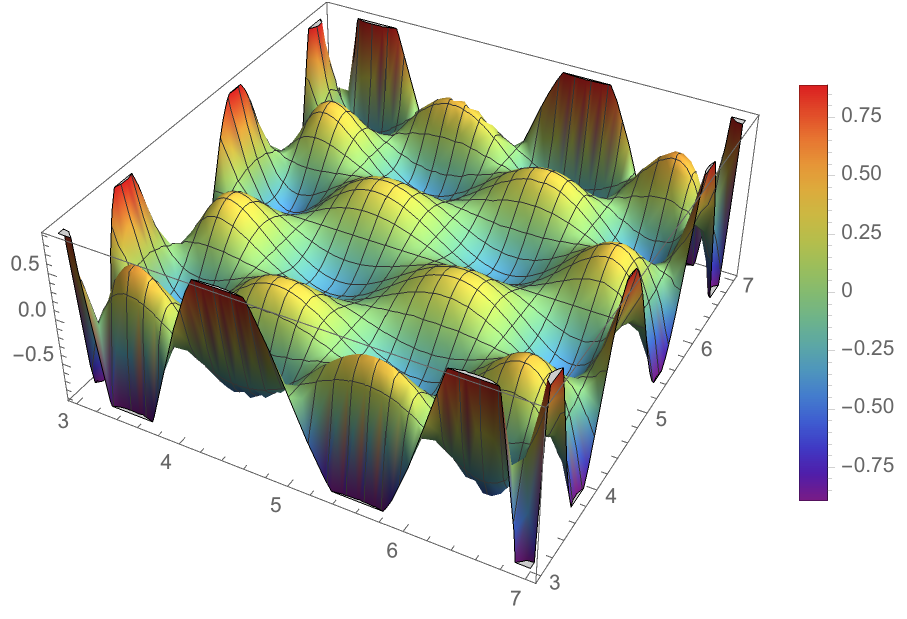}}
\caption{\footnotesize Graphics of  the GLP 
  $W_{\vm}(\va,\vb,\vx)$ 
(\ref{18}, \ref{34}) in 2D  defined in the square $[3,7]\times[3,7]$ for 
$\vm=(3,3)$  and $\vm=(5,7)$, respectively.} 
\label{figure_2}
\end{figure}

This is in, say, two dimensions. The generalization to $n$ dimensions is straightforward.  

We may generalize \eqref{32} as

\begin{equation}\label{36} 
\mathcal R^n\equiv [\va,\vb]^n:= [a_1,b_1]\times  [a_2,b_2]\times \cdots \times [a_n,b_n]\,,\quad 
 \va\,,\, \vb \in \R^n\,,
\end{equation}
and the orthonormal basis \eqref{34} as
\begin{equation}\label{37}
W_{\mathbf m}({\bf a},{\bf b},{\bf x}) \equiv W_{m_1}(a_1,b_1,x_1) \otimes W_{m_2}(a_2,b_2,x_2) \otimes 
  \cdots  \otimes W_{m_n}(a_n,b_n,x_n) \,.
\end{equation}
Now, we have that ${\mathbf m} \in \N^n$,  $\va,\, \vb\in \R^n$ and $\vx$ belongs to $[\va,\vb]^n$. Finally, the orthonormality and completeness relations can be written as, respectively,
\begin{equation}\label{38}
\begin{array}{lll}
\ds\int_{[\va,\vb]^n} d^{\,n}\vx \, W_{\mathbf m'}({\va},{\vb},{\vx})\,
W_{\mathbf m}({\va},{\vb},{\vx})\, =\,\delta_{\mathbf m,\mathbf m'}\,\equiv\, \delta_{m_1,m_1'}\,\delta_{m_2,m_2'}\,\dots\,\delta_{m_n,m_n'}\,,
  \\[3ex]
\ds \sum_{\mathbf m\in \N^{n}}\, W_{\mathbf m}({\va},{\vb},{\vx})\,
W_{\mathbf m}({\va},{\vb},{\vx'})\, =\,\delta (\vx,\vx')\,=\,\delta(x_1-x_1')\,\delta(x_2-x_2')
\,\dots\,\delta(x_n-x_n')\,.
\end{array}
\end{equation}

For every real or complex square integrable function $f(\va,\vb,{\vx}) \in L^2([\va,\vb])$ we have
\begin{equation}\label{39}
f({\va},{\vb,{\vx})\,=\, \sum_{{\mathbf m}={\mathbf 0}}^\infty}  \; f_{[\va,\vb]}^{\mathbf m} \;  W_{\mathbf m}({\va},{\vb},{\vx})
\end{equation}  
where
$f_{[\va,\vb]}^{\mathbf m}$ are the components of  $f({\va},{\vb},{\vx})$  along  $W_{\mathbf m}({\va},{\vb},{\vx})$  and
\begin{equation}\label{40}
  f_{[\va,\vb]}^{\mathbf m} \;=\; \ds
  \int_{[\va,\vb]^n} d^{\,n}\vx \; f({\va,\vb,\vx})\;  W_{\mathbf m}({\bf a},{\bf b},{\vx})\,.
\end{equation}
Explicitly we can write
\begin{equation}\label{41}
f(\mathbf x) \equiv f(x_1,x_2,\dots, x_n) = 
\sum_{m_1,m_2,\dots,m_n=0}^\infty f^{m_1,m_2\dots,m_n}_{[\va,\vb]}\,\bigotimes_{i=1}^{n} W_{m_i}(a_i,b_i,x_i) \,.
\end{equation} 

\begin{figure}[t]
\centering
 \subfigure [$K_{13}(x_1)\otimes K_{13}(x_2)$ \hskip4.15cm (b)   $W_{13}(3,7,x_1)\otimes W_{13}(3,7,x_2)$
]{\includegraphics[width=0.4\textwidth]{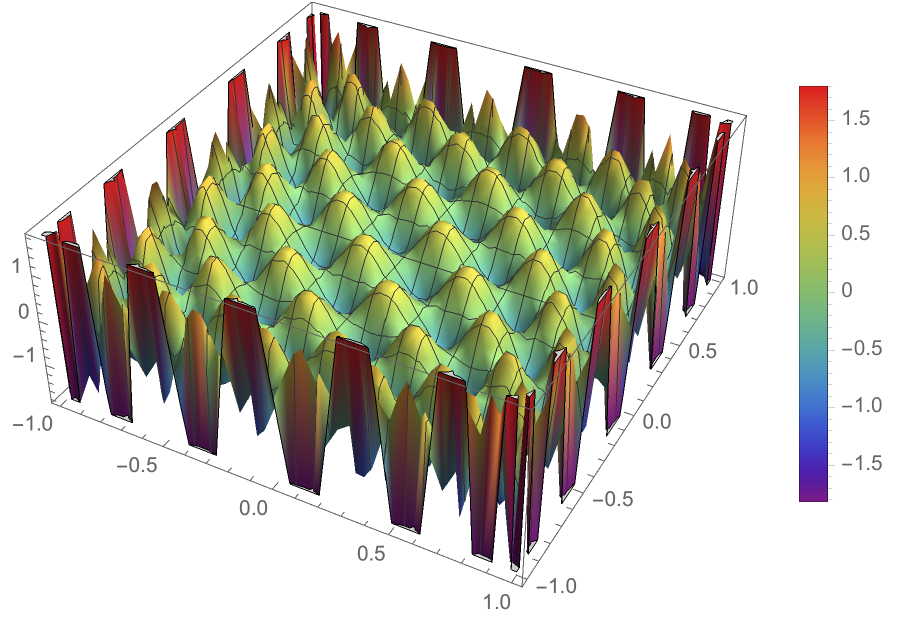}\qquad  
\includegraphics[width=0.4\textwidth]{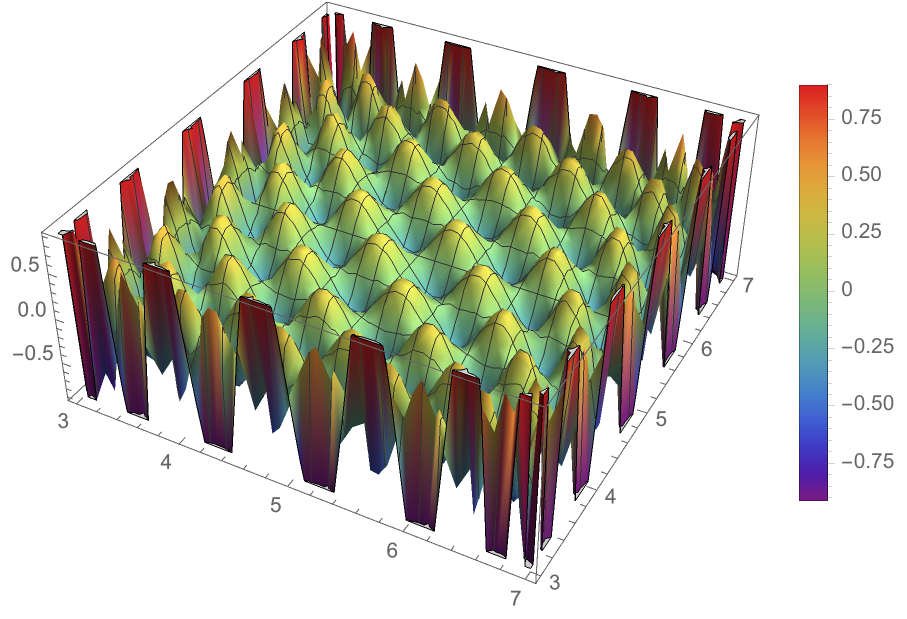}}
\caption{\footnotesize Density plots of: (a)  the NLP  in 2D $K_{\vm}(\vx)$ \eqref{18} in the square $[-1,1]\times[-1,1]$ with $\vm=(13,13)$,  and (b)  the GLP  $W_{\vn}(\va,\vb,\vx)$ 
(\ref{18}, \ref{34})  living in the square $[3,7]\times[3,7]$.}  
\label{figure_3}
\end{figure}

\begin{figure}[t]
\centering
 \subfigure [$W_{13}(2,7,x_1)\otimes W_{12}(2,5,x_2)$ \hskip4.15cm (b)   $W_{13}(2,7,x_1)\otimes W_{12}(2,5,x_2)$]{\includegraphics[width=0.4\textwidth]{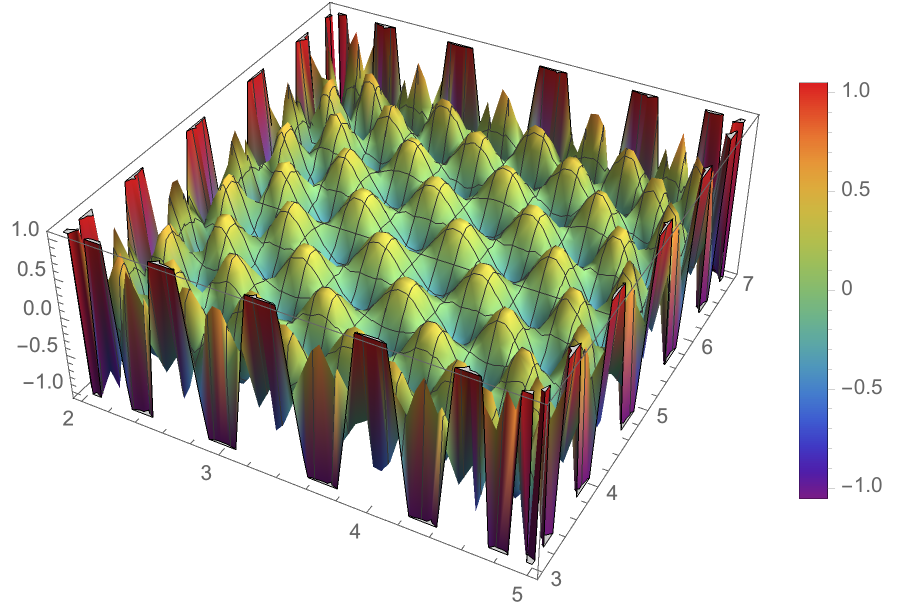}\qquad  
\includegraphics[width=0.4\textwidth]{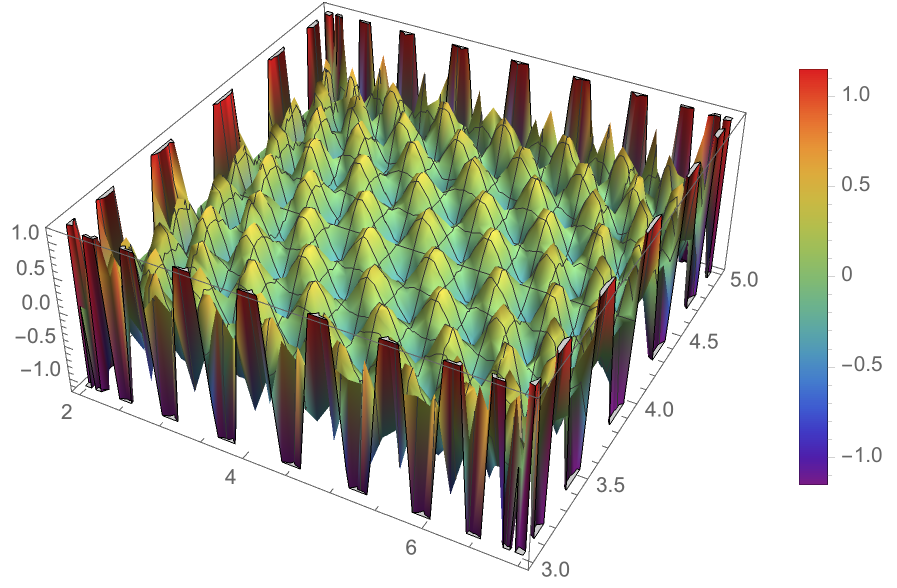}}
\caption{\footnotesize Graphic (a) and Density plot (b) of  the GLP  $W_{\vn}(\va,\vb,\vx)$ 
(\ref{18}, \ref{34}) defined  in  the rectangle $[2,7]\times[2,5]$ and
$\vm=(13,12)$.} 
\label{figure_4}
\end{figure}

So far, the discussion on the more general Legendre polynomials. 


\section{On the smoothness of functions on $\Phi_{a,b}$}

It is clear that the operators $J_\pm$ \eqref{21} and $J_3$ \eqref{28} are unbounded on the Hilbert space $L^2([a,b])$ and, hence, not defined on all vectors of the Hilbert space. Nevertheless,
they have a common subdomain. A proper  locally convex topology on this subdomain, makes the operators $J_\pm$ and $J_3$ continuous. This goes through the construction of a suitable
Gelfand triplet or rigged Hilbert space. Let us start with this construction for $L^2([a,b])$ and then, generalize the procedure to $L^2([\va,\vb]^2)$, which is rather
straightforward. 
Thus, our goal is the construction of a triplet\, $\Phi_{a,b} \subset L^2[(a,b] )\subset \Phi^\times_{a,b}$,\, for which we just need to determine the space
$\Phi_{a,b}$. Let $f(x)$ be an arbitrary function in $L^2([a,b])$, we know that this function can be spanned as
\begin{equation}\label{42}
f(x)= \sum_{m=0}^\infty f^m \, W_m(a,b,x)\,, \,\quad {\rm with} \,\quad \sum_{m=0}^\infty \vert f^m\vert^2 = ||f(x)||^2\,.
\end{equation}
Let us choose $\Phi_{a,b}$ as the space  of functions $f(x)\in L^2([a,b])$, as given in  \eqref{42}, for which
\begin{equation}\label{43}
[p_k(f)]^2:= \sum_{m=1}^\infty |f^m|^2 \, (m+1)^{2k}  < \infty\,,\,\qquad k=0,1,2,\dots\,.
\end{equation}
Note that, for $k=0$, we have the Hilbert space norm. This is a sequence of seminorms, in fact norms, that produce a locally convex topology on $\Phi_{a,b}$. Since this sequence
of seminorms, contains the Hilbert space norm, then, the topology on $\Phi_{a,b}$ is finer than the Hilbert space norm topology. Further considerations show that $\Phi_{a,b}$
with this topology is a complete metrizable space, and hence Fr\`echet \cite{simon1980}. 

The series in \eqref{42} converge in the Hilbert space norm sense. Let see that for  $f(x) \in \Phi_{a,b}$, the series \eqref{42} converge uniformly and absolutely. First of all, from the properties of the Legendre polynomials and \eqref{19}, we have that
\begin{equation}\label{44}
|W_m(a,b,x)| \le \sqrt{\frac{2}{b-a}}\,.
\end{equation}
Then, for any $f(x) \in \Phi_{a,b}$ with span as in \eqref{42}, we have that
\begin{eqnarray}\label{45}
  \left|  \sum_{m=1}^\infty f^m\, W_m(a,b,x)   \right| \le \sum_{m=1}^\infty |f^m|\, |W_m(a,b,x)| \le \sqrt{\frac{2}{b-a}} \sum_{m=1}^\infty |f^m|\, \frac{m+1}{m+1}. \nonumber \\[2ex]
  \le \sqrt{\frac{2}{b-a}} \;\sqrt{\sum_{m=1}^\infty |f^m|^2 (m+1)^2} \; \sqrt{\sum_{m=1}^\infty \frac{1}{(m+1)^2}}\,,
\end{eqnarray}
where the last inequality is just the Cauchy-Schwartz inequality. All the series converge in this last term.  Then, after the Weierstrass M-Theorem, the considered series
$\sum_{m=1}^\infty f^m\, W_m(a,b,x) \in \Phi_{a,b}$ converge absolutely and uniformly. The identity in \eqref{42} for $f(x)\in \Phi_{a,b}$ is also {\it pointwise}.

Before proceeding with our presentation, let us give an important result: let $A$ be a linear operator on a locally convex space $\Phi$, for which the topology is defined by a
set of seminorms $(p_i)_{i\in I}$. Then, $A$ is continuous on $\Phi$ if and only if for each seminorm $p_i$, there exist a positive number $K_i$ and a {\it finite} set of seminorms
$\{p_{i_1}, p_{i_2}, \dots,p_{i_k}\}$, depending on $p_i$, such that \cite{simon1980}
\begin{equation}\label{46}
p_i(A \, f) \le K_i\{ p_{i_1}(f) + p_{i_2}(f) + \dots + p_{i_k}(f)\}\,, \qquad \forall\, f\in \Phi\,.
\end{equation}
This result will be applied in the sequel in order to prove the continuity of the generators of $su(1,1)$ and other linear transformations that we introduce in the sequel.


\subsection{Smoothness}

In the present subsection, we introduce a derivation operation. The idea is as follows:
let $f(x) \in \Phi_{a,b}$ with the form \eqref{42} and satisfying the property \eqref{43}. Let us consider 
\begin{equation}\label{47}
D_xf(x) := \sum_{m=1}^\infty f^m\, W'_m(a,b,x)\,,
\end{equation}
since $W'_0(a,b,x)=0$, 
and  where the prime represents derivation with respect to the variable $x$. We intend to show that $D_x f(x) \in  \Phi_{a,b}$ if $f(x) \in \Phi_{a,b}$ and that the derivation $D_x$ is continuous on $\Phi_{a,b}$. In order to simplify the notation,   we call 
\begin{equation}\label{48}
z:= \frac{2x}{b-a}- \frac{b+a}{b-a}\,,
\end{equation}
hence we have that from \eqref{18} that 
\begin{equation}\label{49}
W'_n(a,b,x) = \left( \frac{2}{b-a}  \right)^{3/2} \, \sqrt{n+1/2} \; P'_n(z) \,.
\end{equation}
Taking into account that  the Legendre  polynomials $P_n(x)$ verify \cite{weberarfken2004}
\begin{equation}\label{50}
\frac{d}{dx}\, P_n(x) = (2(n-1) +1)\, P_{n-1}(x) + (2(n-3) +1) \,  P_{n-3}(x) + \dots \,=\sum_{k=0}^{\lfloor \frac{n-1}{2}\rfloor}  (2(n-1-2 k)+1)\, P_{n-1-2k}(x)\,,
\end{equation}
for $n\geq 1$ (since $d P_0(x)/dx=0$) and where $\lfloor a\rfloor$ stands for the integer part of $a$,
we get that
\begin{equation}\begin{array}{lll}\label{51}
W'_n(a,b,x) &=&\ds \left( \frac{2}{b-a}  \right)^{3/2} \,  \sqrt{n+1/2}\,\sum_{k=0}^{\lfloor \frac{n-1}{2}\rfloor}  (2(n-1-2 k)+1)\, P_{n-1-2k}(z)\,,
 \\ [3ex] 
&=&\ds \left( \frac{2}{b-a}  \right)^{3/2} \,  \sqrt{n+1/2} \,\sum_{k=0}^{\lfloor \frac{n-1}{2}\rfloor}   \frac{\sqrt{(n-1-2k) + 1/2}}{\sqrt{(n-1-2 k) + 1/2}}  \,(2(n-1-2 k)+1)\, P_{n-1-2k}(z)\,,\\[3ex]
& =&   \ds \frac{2}{b-a} \,  \sqrt{n+1/2} \,\sum_{k=0}^{\lfloor \frac{n-1}{2}\rfloor}   \frac{(2n-4 k-1)}{ \sqrt{n-2k-1/2}} \, W_{n-1-2k}(a,b,x)\,,
\qquad n\geq 1 \,.\end{array}\end{equation}
Then, 
\begin{equation}\label{52}
\begin{array}{lll}
D_x f(x)&=& \ds\sum_{m=0}^{\infty} f^{m+1}\, W'_{m+1}(a,b,x) \\[0.4cm]
&=&\ds  \frac{2}{b-a}\, \sum_{m=0}^{\infty}\,\sum_{k=0}^{\lfloor \frac{m}{2}\rfloor} f^{m+1} \,\sqrt{m+3/2}\, 
\frac{(2m-4 k+1)}{ \sqrt{m-2k+1/2}} \, W_{m-2k}(a,b,x)\,.
 \end{array}\end{equation}
Considering $l=m-2k$ (obviously $m\geq 2 k$) and $k=p$ the expression \eqref{52} becomes 
\begin{equation}\label{53}
\begin{array}{lll}
D_x f(x)&=&\ds  \frac{2}{b-a}\, \sum_{l=0}^{\infty}\,\underbrace{2\sqrt{l+1/2}\,\left(\sum_{p=0}^{\infty} f^{l+2 p+1} \,\sqrt{l+2p+3/2}
\right)}_{:=D^l}\,W_{l}(a,b,x)= \frac{2}{b-a}\, \sum_{l=0}^{\infty} D^l\,W_{l}(a,b,x)\,.
 \end{array}\end{equation}

Let us see that the series in \eqref{53} converges absolutely for any arbitrary $l\in \N$. In fact, if we use the Cauchy-Schwartz inequality, we have that
\begin{equation}\begin{array}{lll}\label{54}
\ds \sum_{p=0}^\infty \sqrt{(l+2p) +3/2} \, |f^{l+p+1}| = \sum_{p=0}^\infty  \sqrt{l+2p +3/2} \, \vert f^{l+p+1}\vert\, \
\frac{ (l+p+2)}{(l+p+2)}   \\ [3ex] 
 \hskip2cm\ds \le \left[\sum_{p=0}^\infty (l+2p+3/2) (l+p+2)^2 \, |f^{l+p+1}|^2 \right]^{1/2}  \left[ \sum_{p=0}^\infty \frac{1}{(l+p+2)^2} \right]^{1/2} \,.
\end{array}\end{equation}
Taking into account the following identity, which can be trivially proven,
\begin{equation}\label{55}
(l+2p+3/2) \le (l+p+2)^2\,,
\end{equation}
we have that
\begin{equation}\label{56}
\sum_{p=0}^\infty (l+2p+3/2) (l+p+2)^2 \, \big| f^{l+p+1}\big|^2 \le \sum_{p=0}^\infty (l+p+2)^4 \, \big| f^{l+p+1}\big|^2 \le \sum_{p=0}^\infty (p+1)^4 \, |f^p|^2 = p_2^2(f)\,.
\end{equation}
Also,
\begin{equation}\label{57}
 \sum_{p=0}^\infty \frac{1}{(l+p+2)^2} \le  \sum_{p=0}^\infty \frac{1}{(p+1)^2} =:C^2 <\infty \,,
\end{equation}
which shows that all series in \eqref{54} converge. 

If we consider the norms \eqref{43} we have for \eqref{53} that 
\begin{equation}\label{58}
[p_k(D_x\,f)]^2 =  \left(\frac{2}{b-a}\right)^2  \sum_{l=0} ^\infty \big| D^l\big |^2 (l+1)^{2k}\,.
\end{equation}
Taking into account that the definition of $D^l$ in \eqref{53},  \eqref{56} and \eqref{57}
we have that
\begin{equation}\label{59}
\big| D^l\big| ^2 \le C^2 \, (l+1/2) \sum_{p=0}^\infty (l+p+2)^4 \, |f^{l+p+1}|^2  \le C^2 \, \sum_{p=0}^\infty (l+p+2)^6 \, |f^{l+p+1}|^2 \,,
\end{equation} 
so that, we have the following expression, $k=0,1,2,\dots$ from \eqref{58}
\begin{equation} \begin{array}{lll}\label{60}
\ds \sum_{l=0}^\infty |D^l|^2 \, (l+1)^{2k} & =&\ds  C^2\sum_{l=0}^\infty (l+1)^{2k} \sum_{p=0}^\infty (l+p+2)^6\, |f^{l+p+1}|^2  \\ [3ex] 
&\le&\ds C^2 \sum_{l,p=0}^\infty (l+p+2)^{2(k+3)} \, |f^{l+p+1}|^2\,.
\end{array}\end{equation} 
Now let us consider $s=l+p$. Since for each value of $s$ there are $s+1$ values of $m$ and $p$ such that $s=m+p$ we can rewrite  the last term in \eqref{15} as  
\begin{equation}\label{61}
C^2 \sum_{s=0}^\infty (s+1)(s+1)^{2(k+3)} \, |f^s|^2 \le C^2  \sum_{s=0}^\infty  (s+1)^{2(k+4)} \, |f^s|^2 \le C^2 p^2_{k+4}(f)\,,
\qquad \forall f(x) \in \Phi_{a,b}\,.
\end{equation} 
From the expressions (\ref{58}-\ref{61}), we see that the series in \eqref{58} converges for all values of $k=0,1,2,3,\dots$, so that $D_x f(x)\in \Phi_{a,b}$ if $f(x) \in \Phi_{a,b}$. In addition from \eqref{58} we have 
\begin{equation}\label{62}
[p_k(D_xf)]^2 \le \left(\frac{2}{b-a}\right)^2 C^2 \, p^2_{k+4}(f)\,, \qquad \forall\, f \in \Phi_{a,b}\,,
\end{equation} 
which implies that the derivative in \eqref{53} is a linear continuous mapping on $ \Phi_{a,b}$. From here, we obtain that the {\it same property is valid for the derivatives at all orders}, i.e., functions $f(x) \in \Phi_{a,b}$ are smooth. This property can be extended to $N$ dimensions in the sense that functions as in \eqref{118}  and \eqref{119}  and their generalizations to $n$ dimensions have continuous partial derivatives at all orders on the spaces $\boldsymbol\Phi_{\bf a,b}$ (see end of Section VII).

Question: The derivative $D_x$ in \eqref{47} is the usual derivative? Obviously if the sum in \eqref{47} is finite, this is the usual derivative. Same if the sum is infinite as we have proven that all series of the form $\sum_{m=1}^\infty f^m\, W_m(a,b,x) \in \Phi_{a,b}$ converge uniformly on $[a,b]$.

\subsection{On the multiplication operator and the kets $|a,b,x\rangle$}

Next, let us define a sort of multiplication operator on the space $\Phi_{a,b}$. 
Let us analyze the situation in one dimension first. To begin with, let us consider the following identity between Legendre polynomials 
\cite{weberarfken2004}
\begin{equation}\label{64}
(n+1)P_{n+1}(x) = (2n+1)x P_n(x) - n P_{n-1}(x)\,, \qquad \forall n\in \N.
\end{equation}
Then, using \eqref{48}  and \eqref{18}  and after some cumbersome but otherwise trivial algebra, we obtain the following expression:
\begin{equation}\label{65}
x \,W_m(a,b,x) = \alpha\, W_m(a,b,x) +\beta_{m}\,W_{m+1}(a,b,x)+\gamma_{m}\,W_{m-1}(a,b,x)
\end{equation}
where
\begin{equation}\label{66}
\alpha=\frac{b+a}{2}\,,\qquad
\beta_{m}=\frac{b-a}{2}\,\frac{m+1}{2m+1}\,\sqrt{\frac{m+1/2}{m+3/2}}  \,,\qquad
\gamma_{m}=\frac{b-a}{2}\,\frac{m}{2m+1}\,\sqrt{\frac{m+1/2}{m-1/2}}  \,,
\end{equation}
Thus, for all $f(x) =\sum_{m=0}^\infty f^m\, W_m(a,b,x) \in \Phi_{a,b}$, we define a multiplication operator $Q$ on any $f(x) \in \Phi_{a,b}$ as
\begin{equation}\label{67}
Qf(x) := \sum_{m=0}^\infty f^m\, [x\,W_m(a,b,x)]= \sum_{m=0}^\infty K^m\,W_m(a,b,x)\,.
\end{equation}
According to expressions \eqref{65} and \eqref{66} the coefficient $ K^m$ of $W_m(a,b,x)$ in  \eqref{67}  is given by  three terms
\begin{equation}\label{68}
K^m = \alpha\, f^m + \beta_{m-1} \, f^{m-1} + \gamma_{m+1} \, f^{m+1}\,,
\end{equation}
Note that these three coefficients are bounded by a unique constant, say $c/\sqrt 3$, for all value of $m=0,1,2,\dots$. Using the Cauchy-Schwartz inequality, we have that
\begin{equation}\label{69}
|K^m| \le \sqrt{\alpha^2 + \beta_{m-1}^2 + \gamma_{m+1}^2}\,  \sqrt{|f^m|^2 + |f^{m+1}|^2 + |f^{m-1}|^2} 
\le c \,  \sqrt{|f^m|^2 + |f^{m-1}|^2 + |f^{m+1}|^2}   \,.
\end{equation}

Next, let us consider the following series for $k=0,1,2,\dots$ :
\begin{equation}\label{70}
\sum_{m=0}^\infty |K^m|^2 (m+1)^{2k} \le c \sum_{m=0}^\infty |f^m|^2 (m+1)^{2k} + c \sum_{m=0}^\infty |f^{m-1}|^2 (m+1)^{2k} + c \sum_{m=1}^\infty |f^{m+1}|^2 (m+1)^{2k}\,.
\end{equation}

Note that for the series in \eqref{70}, starting with   the second one and making  $p=m-1$ we have
\begin{equation}\label{71}
\sum_{m=1}^\infty |f^{m-1}|^2 (m+1)^{2k} \le \sum_{p=0}^\infty |f^p|^2 (p+2)^{2k}
\le \sum_{p=0}^\infty |f^p|^2 \left[2(p+1)\right]^{2k}\le 2^{2k}\sum_{p=0}^\infty |f^p|^2\, (p+1)^{2k}\,,
\end{equation}
and analogously 
we have for the third series after the change $p=m+1$
\begin{equation}\label{72}
\sum_{m=0}^\infty |f^{m+1}|^2 (m+1)^{2k} = \le \sum_{p=1}^\infty |f^{p}|^2 (p)^{2k} \le \sum_{p=0}^\infty |f^p|^2 (p+1)^{2k}\,.
\end{equation} 
Hence
\begin{equation}\label{73}
[p_k(Qf)]^2 =\sum_{m=0}^\infty |K^m|^2 (m+1)^{2k} \le 3c\,2^{2k}\, \sum_{m=0}^\infty |f^m|^2 (m+1)^{2k} = 3c \, 2^{2k}\, p_k(f)\,,
\quad
\forall f\in \Phi_{a,b}\,.
\end{equation}
This means that $Qf \in \Phi_{a,b}$, $\forall\, f \in \Phi_{a,b}$ and that $Q$ is continuous on $\Phi_{a,b}$. The same happens to any positive power of $Q$ (i.e., $Q^n$, $n=1,2,3,\dots$). Also $Q$ may be extended with continuity to the dual $\Phi_{a,b}^\times$ by duality. 

This idea may be extended to $n$ dimensions following the definitions given in Section~\ref{nlegendre}. For instance, if $f(\bf x)$ is as in \eqref{41}  with the property 
\begin{equation}\label{74}
[p_{k_1,k_2,\dots, k_n}(f)]^2:= \sum_{{m_1},{m_2,\dots,m_n}=0}^\infty \big|f^{m_1,m_2,\dots,m_n}_{[\va,\vb]}\big|^2\, 
\prod_{i=1}^{n} (m_i+1)^{2k_i} <\infty\,, \quad k_1,k_2,\dots, k_n=0,1,2,\dots\,.
\end{equation}
 we may define the following multiplication operators $\mathbf X_{(l)}=1\otimes 1\otimes \cdots \otimes 1\otimes X\otimes 1\cdots \otimes 1$ where $X$ holds the $l$-th position with $1\leq l\leq n $
\begin{equation}\label{75}
\mathbf X_{(l)}\, f(\mathbf x)  = 
\sum_{m_1,m_2,\dots,m_n=0}^\infty f^{m_1,m_2\dots,m_n}_{[\va,\vb]}\,x_l\,\bigotimes_{i=1}^{n} W_{m_i}(a_i,b_i,x_i) \,.
\end{equation} 
Both,  all their positive powers and their products are continuous operators on $\boldsymbol\Phi_{\bf a,b}$ and can be continuously extended by duality to the dual.


\subsubsection{Continuous and discrete basis}

Let $\mathcal H$ an abstract separable infinite dimensional Hilbert space and let us choose an orthonormal basis $\{|m\rangle\}_{m\in \mathbb N \cup \{0\}}$. For all $f(a,b,x) \in L^2[a,b]$, $\ds f(a,b,x) =\sum_{m=0}^\infty f^m\, W_m(a,b,x)$ the mapping $U: L^2[a,b] \longmapsto \mathcal H$ defined by
\begin{equation}\label{76}
f(a,b,x)\;\;\xrightarrow{U} \;\;|f\rangle := U(f(a,b,x) ) = \sum_{m=0}^\infty f^m\, |m\rangle\, 
\end{equation}
is obviously unitary. Let $\Phi:= U\Phi_{a,b}$ and endow $\Phi$ with the topology transported by $U$ from $\Phi_{a,b}$. This means that the seminorms $q(-)$ on $\Phi$ are of the form $q_k(Uf):= p_k (f) $. Then, we can construct a new RHS 
$(\Phi \subset  \mathcal H \subset  \Phi^\times )$, unitarily equivalent to the former $(\Phi_{a,b}\subset  L^2[a,b] \subset \Phi^\times_{a,b})$ in the following way:
\begin{equation}\label{77}
\begin{array}{lllclll}
&\Phi_{a,b}&\subset & L^2[a,b] &\subset&\Phi^\times_{a,b}
\\[0.2cm]
  & \hskip-0.35cm  {U}\downarrow && \hskip-0.30cm {U}\downarrow &&\hskip-0.30cm {U}\downarrow
\\[0.2cm]
&\Phi &\subset  & \mathcal H &\subset   &\Phi^\times
\end{array}\,,
\end{equation}
where the extension of $U$ into the duals can be made by using duality, i.e., 
\begin{equation}\label{78}
\langle U^{-1} f | F \rangle := \langle f| UF \rangle\,,\qquad \forall f \in \Phi\,,\;\;\forall F \in \Phi_{a,b}^\times\,.
\end{equation}
 This extension $U: \Phi_{a,b}^\times \longmapsto \Phi^\times$ is one to one and onto and also transports any topology compatible with duality (strong, weak, Mackey \cite{HOR}). 

For (almost) all $x\in [a,b]$ with respect to the Lebesgue measure, one may define the functional $|a,b,x\rangle$ by the action on $f\in \Phi$ as
\begin{equation}\label{79}
\langle f | a,b,x\rangle := (U^{-1}f)^*(a,b,x) =: f^*(a,b,x)\,,
\end{equation}
where the star denotes complex conjugation. Note that
\begin{equation}\label{80}
\langle a,b,x|f \rangle := \langle f | a,b,x\rangle^* = f(a,b,x)\,.
\end{equation}
This functional is obviously linear. To show the continuity, note that after \eqref{45}, we have
\begin{equation}\label{81}
|\langle f | a,b,x\rangle| = \left| \sum_{m=0}^\infty f^m \, W_m(a,b,x) \right| \le \sqrt{\frac{2}{b-a}} \, \sqrt{\sum_{m=0}^\infty \frac{1}{(m+1)^2}}\, p_1(U^{-1} f) \equiv C\, q_1(f)\,,
\end{equation}
where the value of the constant $C$ in \eqref{81} is obvious and $ p_1(U^{-1} f)  = q_1(f)$.

The unitary operator $U$ acts on the  functions $f(a,b,x) \in \Phi_{a,b}$ as
\begin{equation}\label{82}
(Q f)(a,b,x) \longmapsto UQ f(a,b,x) = UQU^{-1} f = \widetilde Q f\,,
\end{equation}
where $Q$ is the multiplication operator \eqref{67} and $f = U f(a,b,x)$. Furthermore, for all $f\in \Phi$, we have
\begin{equation}\label{83}
x\, f(a,b,x) = x \langle a,b,x| f \rangle =(Q f)(a,b,x) =  \langle a,b,x| \widetilde Q f \rangle \,.
\end{equation}
Consequently, 
\begin{equation}\label{84}
x \langle a,b,x| =  \langle a,b,x| \widetilde Q\;\; \Longleftrightarrow\;\; \widetilde Q |a,b,x\rangle = x \, |a,b,x\rangle \,,
\end{equation}
so that each of the functionals $|a,b,x \rangle \in \Phi^\times$ is an eigenfunctional of the multiplication operator with eigenvalue $x\in [a,b]$. 

From the above discussion, it is obvious that $|m\rangle \in \Phi$, $m=0,1,2,\dots$. This allows us to give a second interpretation to the identity on $\mathcal H$ given as
\begin{equation}\label{85}
I := \sum_{m=0}^\infty |m\rangle \langle m| \,.
\end{equation}
According to definitions \eqref{79} and \eqref{80}, we have that $\langle m| a,b,x \rangle = W_m(a,b,x)$, so that
\begin{equation}\label{86}
|a,b,x\rangle = \sum_{m=0}^\infty |m\rangle \langle m| a,b,x\rangle = \sum_{m=0}^\infty W_m(a,b,x)\, |m\rangle\,.
\end{equation}
 If $|F\rangle$ were an arbitrary functional in $\Phi^\times$, we would have
\begin{equation}\label{87}
|F\rangle = \sum_{m=0}^\infty |m\rangle \langle m|F\rangle\,,
\end{equation}
so that $I$ as in \eqref{85} can be looked at an identity on $\Phi^\times$ and this identity provides a nice series representation of each functional in $\Phi^\times$. This series converges in the sense that 
\begin{equation}\label{88}
\langle f|F\rangle= \sum_{m=0}^\infty \langle f|m\rangle \langle m |F \rangle\,,\qquad \forall f\in \Phi\,.
\end{equation}

Unitary mappings preserve the scalar product, so that if $f,g \in \Phi$ and $f(a,b,x) = U^{-1}f$ and $g(a,b,x) = U^{-1}g$, one has
\begin{equation}\label{89}
\langle f| g \rangle = \int_a^b f^*(a,b,x) g(a,b,x)\, dx = \int_a^b \langle f |a,b,x\rangle\langle a,b,x| g \rangle\, dx\,.
\end{equation}
If we omit the arbitrary $f \in \Phi$, we have the following formal expression:
\begin{equation}\label{90}
|g\rangle = \int_a^b \langle a,b,x| g \rangle \, |a,b,x \rangle \, dx\,.
\end{equation}
This formal expression gives an split of each $g \in \Phi$ in terms of the so called {\it continuous basis} $\{ |a,b,x \rangle\}$, where $a$ and $b$ are fixed real numbers with $a<b$ and $x\in [a,b]$. In \eqref{86}, we have written the elements of this continuous basis in terms of the discrete basis (or orthonormal basis) $\{ |m \rangle\}$. Then an inversion formula should be in order. Obviously, this comes directly from \eqref{90}, since for all $m=0,1,2,\dots$, we have
\begin{equation}\label{91}
|m\rangle = \int_a^b \langle a,b, x| m\rangle \, |a,b,x\rangle \, dx = \int_a^b W_m(a,b,x)\, |a,b,x \rangle\, dx \,.
\end{equation}
Also, if we omit the arbitrary $f,g \in \Phi$ in \eqref{89}, we obtain the following formal identity:
\begin{equation}\label{92}
\mathcal I = \int_a^b  |a,b,x\rangle\langle a,b,x|  \, dx \,.
\end{equation}
Since after \eqref{90} $\mathcal I g =g$, this $\mathcal  I$ as in \eqref{92} represents the canonical injection $\mathcal I:\Phi \longmapsto \Phi^\times$. 


\subsubsection{On the essential self-adjointness of $Q$ on $\Phi_{a,b}$.}

After the definition \eqref{68} and its continuity on $\Phi_{a,b}$, one sees that $Q$ is symmetric on $\Phi_{a,b}$.  
Effectively, we have to prove that $\langle Q\,f,g\rangle=\langle f,Q\,g\rangle$ for all $f,g\in\Phi_{a,b}$. Taking into account \eqref{68}, we have that
\begin{equation}\label{93}
\begin{array}{lll}
\langle Q\,f,g\rangle =\ds \int_a^b \left( \sum_{m=0}^\infty K^m\,W_m(a,b,x)\right)^*\, \left(\sum_{n=0}^\infty g^n\,W_n(a,b,x)\right)\,dx
&=&\ds\sum_{m=0}^\infty [K^m]^*\,g^m\\[0.4cm]
&=&\ds\sum_{m=0}^\infty \alpha\, [f^m]^*\,g^m + \sum_{m=1}^\infty \beta_{m-1} \, [f^{m-1}]^* \,g^m +
 \sum_{m=0}^\infty \gamma_{m+1} \, [f^{m+1}]^* g^m\,.
\end{array}\end{equation}
We also have that
\begin{equation}\label{94}
\langle f,Q\,g\rangle=\sum_{m=0}^\infty [f^m]^*\,\left( \alpha\, g^m + \beta_{m-1} \, g^{m-1} + \gamma_{m+1} \, g^{m+1}\right)=
\sum_{m=0}^\infty\,\alpha\, [f^m]^* \, g^m +  \sum_{m=1}^\infty \beta_{m-1} \, [f^m]^* \,g^{m-1} 
+ \sum_{m=0}^\infty \gamma_{m+1} \,[f^m]^* \, g^{m+1}\,.
\end{equation}
From the series of  \eqref{93}  we arrive to 
\begin{equation}\label{95}
 \sum_{m=1}^\infty \beta_{m-1} \, [f^{m-1}]^* g^m \;\;\stackrel{m-1=n}{=}\;\; \sum_{n=0}^\infty \beta_{n} \, [f^{n}]^* g^{n+1}\,,\qquad
\sum_{m=0}^\infty \gamma_{m+1} \, [f^{m+1}]^* g^m \;\;\stackrel{m+1=n}{=}\;\; \sum_{n=1}^\infty \gamma_{n} \, [f^{n}]^* g^{n-1}
 \end{equation}
  Using the definitions \eqref{66} we easily proof that $\beta_{n}=\gamma_{n+1}$ and  $\gamma_{n}=\beta_{m-1}$. Hence we arrive to
 $\langle Q\,f,g\rangle=\langle f,Q\,g\rangle$,   $\forall\, f,g \in \Phi_{a,b}$.

Then, in order to prove the essential self adjointness of $Q$ on $\Phi_{a,b}$, we need to show that the range of $Q \pm iI$, where $I$ is the identity operator, is dense on $L^2[a,b]$. Equivalently, we may show that if there exists a function $f\in L^2[a,b]$ such that for all $m=0,1,2,\dots$
\begin{equation}\label{96}
\langle f | (Q \pm iI)\, W_m \rangle =0\, 
\end{equation}
then, this function is the zero function. Let us show that this is the case with $+i$, the case with $-i$ being identical. For an arbitrary value of $m=0,1,2,\dots$, we have that, after \eqref{65}, we have
\begin{equation}\label{97}
\langle f | (Q +i I)\, W_m \rangle = (\alpha +i) \langle f | W_m \rangle + \beta_{m-1} \langle f | W_{m-1} \rangle + \gamma_{m+1} \langle f |W_{m+1} \rangle =0\,.
\end{equation}
Note that $c_m:= \langle f | W_m \rangle$ is the $m$-th component of $f$ in the orthonormal basis $\{ W_m(a,b,x)\}$, so that $f(x) = \sum_{m=0}^\infty c_m \, W_m(a,b,x)$ in the norm sense. This gives the following sequence of equations ($\beta_{-1}=0$):
\begin{equation}\label{98}
(\alpha +i) c_m + \beta_{m-1} c_{m-1} + \gamma_{m+1}c_{m+1} =0\,, \qquad m=0,1,2,\dots\,.
\end{equation}
It is interesting to see that all $c_m$ may be written in terms of $c_0$. Let us begin with \eqref{98} with $m=0$. It gives:
\begin{equation}\label{99}
(\alpha +i) c_0 + \gamma_1 c_1 =0 \Longrightarrow c_1 = - \frac{\alpha+i}{\gamma_1}\, c_0\,.
\end{equation}
For $m=1$, one has
\begin{equation}\label{100}
(\alpha + i) c_1 + \beta_0 c_0 + \gamma_2 c_2 =0 \Longrightarrow c_2 = -\frac{1}{\gamma_2} \left[ \frac{(\alpha + i)^2}{\gamma_1} + \beta_0 \right] c_0\,.
\end{equation}
Similar equations for $m=3,4$ give, respectively,
\begin{equation}\label{101}
c_3 = \left\{ -\frac{1}{\gamma_3}  (\alpha +i) \frac{1}{\gamma_2} \left[ \frac{(\alpha+i)^2}{\gamma_1} + \beta_0 \right] + \frac{(\alpha +i ) \beta_1}{\gamma_3 \gamma_1} \right\} c_0\,,
\end{equation}
and
\begin{equation}\label{102}
c_4 = \left\{ \frac{1}{\gamma_4} (\alpha+i) \frac{1}{\gamma_3} (\alpha+i) \frac{1}{\gamma_2} \left[ \frac{(\alpha+i)^2}{\gamma_1} + \beta_0 \right] +  \frac{(\alpha +i ) \beta_1}{\gamma_3 \gamma_1} - \frac{\beta_2}{\gamma_4 \gamma_2}  \left[ \frac{(\alpha+i)^2}{\gamma_1} + \beta_0 \right]\right\} c_0 \,.
\end{equation}
We have obtained enough terms to understand the behaviour of $c_m=[-]_mc_0$. In fact, $[-]_m$ is the sum of fractions having the same number of terms in the numerator and the denominator. A term by term analysis shows that $[-]_m$  does not go to zero as $m\longmapsto\infty$. For instance, let us study the first term in $[-]_m$, which has the following form:
\begin{eqnarray}\label{103}
\frac{(\alpha +i)^m}{\gamma_m \gamma_{m-1}\dots\gamma_1}= \left[ \frac{a+b +2i}{2} \right]^m \left( \frac{b-a}{2} \right)^{-m}\, C_m^{-1} C_{m-1}^{-1}  \dots C_1^{-1}\,,
\end{eqnarray}
where
\begin{equation}\label{104}
C_m = \frac{m}{2m+1} \sqrt{\frac{m+1/2}{m-1/2}}\,.
\end{equation}
Observe that the sequence with general term $C_m$ is a decreasing sequence, with $1/2 < C_m \le \sqrt 3/3$. Then, $C_m^{-1} C_{m-1}^{-1}  \dots C_1^{-1} > (3/\sqrt 3)^m$. Also consider
\begin{equation}\label{105}
\left[ \frac{a+b +2i}{2} \right]^m \left( \frac{b-a}{2} \right)^{-m} = \left[ \frac{a+b}{b-a} +\frac{2i}{b-a} \right]^m\,.
\end{equation}
This is a complex number with modulus 
\begin{equation}\label{106}
\left[ \left| \frac{a+b}{b-a} \right|^2 + \frac{4}{(b-a)^2} \right]^{m/2} \,,
\end{equation}
so that if $a$ and $b$ had the same sign, it results that \eqref{106} goes to infinite as $m \longmapsto \infty$. This may not be the case if $b$ were positive and $a$ negative. A similar analysis can be made of all other terms except the last one, which contain a term of the form $(\alpha +i)^p$ in the numerator and a product of $p$ terms of the form $\gamma_i$ (indeed more) in the denominator. Then, if $a$ and $b$ have different sign they may go to zero as $m$ goes to infinite. We even take into account that the number of terms in $c_m$ increase with $m$, the sum may go to zero. So far the situation is not clear. 
The difference is in the last term, which has the form,
\begin{equation}\label{107}
r_m:= \frac{\beta_{2m-2}\dots\beta_0}{\gamma_{2m} \dots \gamma_2}\,.
\end{equation}

Let us show that the quotient $\beta_{2m-2}/\gamma_{2m}>1$, $m=1,2\dots$. Note that all therms of the form $(b-a)/2$ cancel out here. In fact,
\begin{equation}\label{108}
\frac{\beta_{2m-2}}{\gamma_{2m}} = \sqrt{\frac{2m-3/2}{2m+1/2}}\, \frac{2m-1}{4m-3}\, \frac{4m+1}{2m}\,.
\end{equation}
Let us square the right hand side in \eqref{108} and then, multiply by 2 numerator and denominator in the first factor. We obtain,
\begin{equation}\label{109}
\frac{4m-3}{4m+1}\, \frac{(2m-1)^2}{(4m-3)^2}\, \frac{(4m+1)^2}{4m^2} = \frac{(2m-1)^2}{4m^2} \, \frac{4m+1}{4m-3} = \frac ND\,,
\end{equation}
where $N$ and $D$ stands for numerator and denominator, respectively. Note that $N= D+1$, so that $N>D$ and, hence, $\sqrt{N/D}>1$. 
Then, the expression in \eqref{107} is always bigger than one and therefore, the series $\sum_{m=0}^\infty |r_m|^2$ always diverges. The conclusion is that the series $\sum_{m=0}^\infty |c_m|^2$ always diverges, so that it does not exist a function $f(x) \in L^2[a,b]$ verifying \eqref{96} for all $m=0,1,2,\dots$. The essentially self adjointness of $Q$ is, then, proved.

However $D_x$ is not even symmetric. Effectively, from  \eqref{51} and for short $W_m \equiv W_m(a,b,x)$ we can compute,  for instance, 
\begin{equation}\label{110}
\langle W_m | D_x W_{m+1}\rangle = \frac{2}{b-a} \, (2n+1)\, \frac{\sqrt{n+3/2}}{ \sqrt{n+1/2}} \,\langle W_m|W_m \rangle =  \frac{2}{b-a} \, (2n+1) \, \frac{\sqrt{n+3/2}}{ \sqrt{n+1/2}} \ne 0\,.
\end{equation}
On the other hand,
\begin{equation}\label{111}
\langle D_x W_m| W_{m+1} \rangle =0 \Longrightarrow \langle D_x W_m| W_{m+1} \rangle \ne \langle W_m | D_x W_{m+1}\rangle\,.
\end{equation}
The point is that if $f(x) \in \Phi_{a,b}$, in general $f(b)\ne 0$ and $f(a) \ne 0$, so that
\begin{equation}\label{112}
\langle D_x f| g \rangle = \int_a^b f'(x)\, g(x) \, dx = f(b) g(b) - f(a) g(a) - \int _a^b f(x)\, g'(x)\, dx \,.
\end{equation}
Nevertheless, the action of $D_x \pm iI$ on $\Phi_{a,b}$ is dense in $L^2[a,b]$. The proof is a bit cumbersome, although not complicated.


\section{On the continuity of the generators of $su(1,1)$}\label{rhs}

To begin with, we should first define the action of $J_\pm$ on $\Phi_{a,b}$, which could be made after  \eqref{24}. 
 Thus, the action of $J_+$ on $f(x) \in \Phi_{a,b}$ should be defined as 
\begin{equation}\label{113}
J_+ \sum_{m=1}^\infty f^m\, W_m(a,b,x) := \sum_{m=1}^\infty f^m \, (m+1)\, W_{m+1}(a,b,x)\,.
\end{equation}
We need to show that this is a good definition in the sense that the sum converges to an element of $\Phi_{a,b}$ and that $J_+$ is continuous on $\Phi_{a,b}$. The former is just
a trivial consequence of the definition of $\Phi_{a,b}$ (it is obviously linear). To prove the continuity, take any  seminorm $p_k$, then, for any $f(x)\in \Phi_{a,b}$ we have
\begin{eqnarray}\label{114}
  [p_k(J_+f)]^2 &=& \sum_{n=1}^\infty |f^m|^2\, (m+1)^2\, (m+2)^{2k} \le \sum_{m=1}^\infty\, |f^m|^2\, (m+1)^2\, (2m+2)^{2k} \nonumber \\ [2ex]
  &=& 2^{2k} \sum_{m=1}^\infty\, |f^m|^2\, (m+1)^{2(k+1)} = 2^{2k}\, [p_{k+1}f]^2\,, \quad k=0,1,2,\dots\,,
\end{eqnarray}
which, after \eqref{46}, shows the continuity of $J_+$ on $\Phi_{a,b}$. 

The definition of $J_-$ on $\Phi_{a,b}$ should be given by
\begin{equation}\label{115}
J_- \sum_{m=1}^\infty f^m\, W_m(a,b,x) := \sum_{m=2}^\infty f^m\, m\, W_{m-1}(a,b,x)\,.
\end{equation}

The operator $J_-$ is obviously linear and well defined on $\Phi_{a,b}$. Furthermore, for all $ f(x) \in \Phi_{a,b}$
\begin{equation}\label{116}
[p_k(J_- f)]^2 = \sum_{m=2}^\infty |f^m|^2\, m^2\, (m-1+1)^{2k} \le \sum_{m=1}^\infty |f^m|^2\, (m+1)^{2(k+1)}
 = [p_{k+1}f]^2\, , \quad k=0,1,2,\dots\,. 
\end{equation}
This shows the continuity of $J_-$ on $\Phi_{a,b}$.  

Analogously, we define $J_3$ on $\Phi_{a,b}$ as
\begin{equation}\label{117}
J_3  \sum_{m=1}^\infty f^m\, W_m(a,b,x) = \sum_{m=1}^\infty f^m\,\left(m+\frac12\right)\, W_m(a,b,x)\,.
\end{equation}
By repeating the same argument, we show that $J_3$ is well defined and continuous on $\Phi_{a,b}$. Note that all the three operators, as well as their products and linear combinations
may be extended by continuity to the dual $\Phi^\times_{a,b}$, provided that this dual be endowed with any topology compatible with the dual pair, usually the weak
or the strong topologies \cite{HOR}.

These results may be extended to the general situation studied in Section 5. In the simplest case of dimension equal to two, an arbitrary function
$f(\mathbf x) \in L^2([\mathbf a, \mathbf b]^2)$ can be written as \eqref{42}
\begin{equation}\label{118}
f(\mathbf x) \equiv f(x_1,x_2) =  \sum_{\vm={\mathbf 0}}^{\infty}  \; f_{[\va,\vb]}^{\mathbf m} \;  
W_{\mathbf m}({\va},{\vb},{\vx})=
\sum_{m_1,m_2=0}^\infty f^{m_1,m_2}_{[\va,\vb]}\, W_{m_1}(a_1,b_1,x_1) \, W_{m_2}(a_2,b_2,x_2)\,.
\end{equation} 
The space $\Phi_{\mathbf a, \mathbf b}$ is defined as the space of functions $f(x_1,x_2) \in L^2([\mathbf a, \mathbf b])$ such that (see \eqref{39})
\begin{equation}\label{119}
[p_{k_1,k_2}f]^2:= \sum_{{m_1},{m_2}=0}^\infty \big|f^{m_1,m_2}_{[\va,\vb]}\big|^2\, (m_1+1)^{2k_1}\, (m_2+1)^{2k_2} <\infty\,, \quad k_1,k_2=0,1,2,\dots\,.
\end{equation}
All the tensor products of the operators $J_\pm$ and $J_3$ (also their linear combinations and products)  act linearly and are continuous on $\Phi_{\mathbf a, \mathbf b}$. Proofs are
exactly as above for dimension one. The extension of all these ideas to n dimensions is straightforward.


\section{Concluding remarks}\label{remarks}

In this paper we present a generalization of the Legendre polynomials to any finite interval $[a,b]$ of the real line.
These new functions, $W(m,a,b,x)$, constitute an orthonormal and countable basis for the square integrable functions defined in this interval. We use these functions to construct bases in `$n$-rectangular' domains 
$\mathcal R^n=[a_1,b_1]\times[a_2 ,b_2]\times\dots\times[a_n,b_n]\subset \R^n$
as $W(m_1,a_1,b_1,x_1)\otimes W(m_2,a_2,b_2,x_2)\otimes\dots\otimes W(m_n,a_n,b_n,x_n)$.

A practical application of this construction could be the reconstruction of images in a kind of `analogical' procedure  in 2D or 3D either in black and white or in color. To this end, we span some function in terms of series of $W$-functions and just keep a finite number of terms. 
Note that from the enormous contents of information of an image,  we -the humans- can just use a minimal part. The problem is to give a criteria on how to obtain the relevant part of the information.

Using these Legendre polynomials, we construct Gelfand a triplet supporting a representation of $su(1,1)$ by continuous operators on that triplet. We show a property of smoothness on this triplet and the possibility of constructing a multiplication operator on it, which enables us for the construction of so called {\it continuous} bases. Relations between usual {\it discrete} orthonormal basis and continuous bases are well established. All these notions may be extended to $n$-dimensional systems.


\section*{Acknowledgements}

The paper has been partially supported by the  Q-CAYLE project, funded by the European Union-Next Generation UE/MICIU/Plan de Recuperacion, Transformacion y Resiliencia/Junta de Castilla y Leon (PRTRC17.11), and also by RED2022-134301-T, PID2020-113406GB-I00, PID2023-148409NB-I00 and PID2023-149560NB- C21 
financed by MICIU/AEI/10.13039/501100011033.  This article/publication is based upon work from COST Action CaLISTA CA21109 supported by COST (European Cooperation in Science and Technology).


\end{document}